\documentclass[12pt,preprint]{aastex}

\usepackage{amsmath}
\usepackage{amsfonts}
\usepackage{amssymb}
\usepackage{float}
\usepackage{graphicx}
\usepackage{gensymb}

\newcommand\cm{{\rm\,cm}}

\title{Polarized Synchrotron Emissivities and Absorptivities for Relativistic Thermal, Power-Law, and Kappa Distribution Functions}
\author{Alex Pandya}
\affil{Department of Physics, University of Illinois, 1110 West Green Street, Urbana, IL, 61801, {\tt aapandy2@illinois.edu}}
\author{Zhaowei Zhang}
\affil{Department of Physics, University of Illinois, 1110 West Green Street, Urbana, IL, 61801, {\tt zzhan119@illinois.edu}}
\author{Mani Chandra}
\affil{Department of Astronomy, University of Illinois, 1002 West Green Street, Urbana, IL, 61801, {\tt manic@illinois.edu}}
\author{Charles F. Gammie}
\affil{Department of Astronomy, University of Illinois, 1002 West Green Street, Urbana, IL, 61801, {\tt gammie@illinois.edu} }

\begin{document}

\begin{abstract}

Synchrotron emission and absorption determine the observational appearance of many astronomical systems.  In this paper, we describe a numerical scheme for calculating synchrotron emissivities and absorptivities in all four Stokes parameters for arbitrary gyrotropic electron distribution functions, building on earlier work by Leung, Gammie, and Noble.  We use this technique to evaluate the emissivities and the absorptivities for a  thermal (Maxwell-J\"uttner), isotropic power-law, and isotropic kappa distribution function.  The latter  contains a power-law tail at high particle energies that smoothly merges with a thermal core at low energies, as is characteristic of observed particle spectra in collisionless plasmas. We provide fitting formulae and error bounds on the fitting formulae for use in codes that solve the radiative transfer equation. The numerical method and the fitting formulae are implemented in a compact C library called {\tt symphony}.  We find that: the kappa distribution has a source function that is indistinguishable from a thermal spectrum at low frequencies and transitions to the characteristic self-absorbed synchrotron spectrum, $\propto \nu^{5/2}$, at high frequency; the linear polarization fraction for a thermal spectrum is near unity at high frequency; and all distributions produce O(10\%) circular polarization at low frequency for lines of sight sufficiently close to the magnetic field vector.
\end{abstract}

\section{Introduction}

Cyclotron and synchrotron emission (or {\em magnetobremsstrahlung}) is produced by energetic electrons spiraling in a magnetic field in relativistic jets, in accretion flows onto black holes, in the interstellar medium, stellar coronae, planetary magnetospheres, and many other settings.  Our particular interest is in modeling synchrotron emission from potential Event Horizon Telescope targets \citep[e.g.][]{Doeleman2009}, including the galactic center source Sgr A*.  We are especially interested in modeling polarized emission, which may carry information about magnetic field structure in Sgr A* that is unavailable in the total intensity \citep[e.g.][]{Johnson2014}.

The problem of polarized transfer in a magnetized hot plasma is somewhat involved \citep[see, e.g.][for a covariant treatment]{Gammie2012}. As input one must specify the electron distribution function $f$ and the magnetic field strength and direction.  Given these it is possible to calculate the emission and absorption coefficients using a suitable description of the radiation fields (the Stokes parameters are what we use here).  In addition, for the full transfer problem, one must evaluate the Faraday rotation and conversion coefficients.

Evaluations of the synchrotron emission and absorption coefficients date back to at least \cite{Westfold1959}, who used an ultrarelativistic approximation (although one issue that arises in the context of the galactic center is that the electrons are at most only mildly relativistic).  The basic results are summarized in the important review paper of \cite{Ginzburg1965}. More recent work is summarized in \cite{Leung2011} and \cite{Dexter2016}.  

In this paper our goal is to extend the work of \cite{Leung2011} in three directions.  First, one would like to be able to treat polarized radiation rather than just total intensity.  Second, Leung et al. give a fitting formula only for a relativistic thermal distribution of electrons; here we consider thermal, power-law, and the so-called kappa distribution \citep[e.g.][]{Vasyliunas1968}, which has a thermal core and power-law tails.   Third, the Leung et al. code, {\tt harmony}, is not as robust, easy to use, and modify as one might hope. Here we extend the numerical scheme to make it more robust and use it to evaluate synchrotron emissivities and absorptivities for all Stokes parameters. The scheme allows the computation to be performed for arbitrary gyrotropic distribution functions. We implement the scheme in a new code {\tt symphony}, as well as include fits specialized to the above three distribution functions in the same code with a consistent interface. For example, to evaluate the emissivities directly for any distribution function, one uses the {\tt j\_nu()} function, and to evaluate the fits to the emissivities for the above three distribution functions, one can call the {\tt j\_nu\_fit()} function. The numerical scheme and the interface in {\tt symphony} makes it easy for one to incorporate new models, such as anisotropic distribution functions.

To accomplish the three extensions of Leung et al. listed above, we begin in \S \ref{sec:radtrans} 
by setting notation, reviewing the polarized transfer equation, and writing the expressions that must be evaluated to find the emissivities and absorptivities.  In \S \ref{sec:DF} we review the model electron distribution functions; in \S \ref{sec:methods} we summarize the necessary numerical procedure for evaluating what is essentially a two-dimensional integral and test our numerical procedure.  The key results are in \S \ref{sec:results}, which compares emissivities and source functions from the three distribution functions, shows emitted polarization fractions, and provides fitting formulae.  Then \S \ref{sec:conclusion} summarizes the results and gives a guide to the code.

\section{Radiative Transfer}\label{sec:radtrans}

We are interested in incoherent but polarized radiation, which can be described in the Stokes basis $I_S = \{I, Q, U, V\}$. Recall that $I$, also written $I_\nu$ ($\nu \equiv$ frequency)  is the total intensity, $Q$ and $U$ describe linear polarization, and $V$ measures circular polarization.  

In what follows we use notation consistent with \cite{Leung2011}: $\nu$ is the frequency and ${\textbf{k}}$ is the wavevector of the emitted or absorbed photon, which lies at an angle $\theta$ to the magnetic field vector~$\textbf{B}$ (unless otherwise stated $B$ is measured in Gauss); $\textbf{p} = m_e \gamma \textbf{v}$ is the electron momentum and $\textbf{v}$ is the electron velocity, which lie at pitch angle $\xi$ to $\textbf{B}$; here $\gamma \equiv$ Lorentz factor and $\beta \equiv v/c$.  Finally, $\nu_c = e B/(2 \pi m_e c) = 2.8 \times 10^6 \, B \, {\rm Hz} \equiv$ electron cyclotron frequency.  

We orient coordinates in the plasma frame so that $Q > 0$ for polarization vectors in the $\textbf{k}$-$\textbf{B}$ plane and as usual $Q < 0$ for polarization perpendicular to that plane.  Since synchrotron emission is linearly polarized perpendicular to the wavevector-magnetic field plane, $j_Q < 0$ and $j_U = 0$, by symmetry.  Consistent with IEEE and IAU conventions, $V > 0$ if the electric field vector rotates in a right-handed sense around the wavevector.

We assume that the distribution function is {\em gyrotropic}, i.e. independent of gyrophase, consistent with the idea that, in many applications, the electron Larmor radius $r_g = 1.7 \times 10^3 \beta\gamma/B \cm$ is small compared to the size of the system and small compared to characteristic scales in the flow.  In writing the emissivities and absorptivities below we also assume that $(\nu/\nu_p)^2 \gg 1$ and that $(\nu/\nu_p)^2 (\nu/\nu_c) \gg 1$; here $\nu_p \equiv (n_e e^2/(\pi m_e))^{1/2} = 8980 n_e^{1/2} {\rm Hz}$ is the plasma frequency.  

In the Stokes basis the radiative transfer equation is 
\begin{equation}
\frac{d}{ds}I_{S}  = J_{S} - \textbf{M}_{ST}I_{T}
\end{equation}
where the vector $J_{S} = \{j_{I}, j_{Q}, j_{U},j_{V}\}^{T}$ contains the emission coefficients, which have units of $dE/dtdVd{\nu}d\Omega$ ($V \equiv$ volume, $d\Omega \equiv$ differential solid angle).  The Mueller matrix $M_{ST}$ is
\begin{equation}
M_{ST} = 
\begin{pmatrix}
\alpha_{I}    &\alpha_{Q}    &\alpha_{U}    &\alpha_{V}    \\
\alpha_{Q}    &\alpha_{I}    &r_{V}         &-r_{U}       \\
\alpha_{U}    &-r_{V}        &\alpha_{I}    &r_{Q}         \\
\alpha_{V}    &r_{U}         &-r_{Q}        &\alpha_{I}    
\end{pmatrix}
\end{equation}
where $\alpha_{S}$ are the absorption coefficients and $r_{Q}$, $r_{U}$, and $r_{V}$ are what we will call Faraday mixing coefficients.  Approximate expressions for these coefficients are given in \cite{huang2011} and \cite{Dexter2016}.

In this paper we calculate $\alpha_S$ and $j_S$ for several distribution functions.  As described (but not originally) by \cite{Leung2011}, the emissivity in the Stokes basis is related to the distribution function by 
\begin{equation} \label{eq:EmiStokes}
J_S = \frac{2 \pi e^2 \nu^2}{c} \int d^3 p \,\, f \,\, \sum_{n=1}^{\infty} \delta(y_n) K_S, 
\end{equation}
where 
\begin{equation}\label{eq:KIdef}
K_I = M^{2}J^{2}_{n}(z) + N^{2}J^{\prime2}_{n}(z), 
\end{equation}
\begin{equation}\label{eq:KQdef}
K_Q = M^{2}J^{2}_{n}(z) - N^{2}J^{\prime2}_{n}(z), 
\end{equation}
\begin{equation}
K_U = 0,
\end{equation}
and
\begin{equation}\label{eq:KVdef}
K_V = -2MNJ_{n}(z)J_{n}^{\prime}(z).
\end{equation}
Here
\begin{equation}
M = \frac{\cos\theta-\beta \cos\xi}{\sin\theta}
\end{equation}
\begin{equation}\label{eq:Ndef}
N = \beta \sin\xi
\end{equation}
\begin{equation}
y_n = \frac{n \nu_c}{\gamma} - \nu(1 - \beta \cos\xi \cos\theta)
\end{equation}
and
\begin{equation}\label{eq:zdef}
z = \frac{\nu \gamma \beta \sin\theta \sin\xi}{\nu_c}.
\end{equation}
Here $J_n$ is a Bessel function of the first kind, and $J_n'$ is its derivative.

The absorptivities in the Stokes basis are 
\begin{equation} \label{eq:AbsStokes}
\alpha_{S} = -\frac{ce^2}{2\nu}\int d^{3}p  \,\, Df \,\, \sum_{n=1}^{\infty}\delta(y_{n})  K_{S}.
\end{equation}
where (assuming a gyrotropic distribution function)
\begin{equation}
Df \equiv \frac{2\pi\nu}{m_e c^2}\left(\frac{\partial}{\partial\gamma} + \frac{\beta\cos{\theta} - \cos\xi}{\beta^2 \gamma}\frac{\partial}{\partial\cos\xi}\right)f.
\end{equation}
We have treated the electron classically so that in the absorptivity, for example, the change in electron momentum associated with absorption of a single photon is small compared to the width of the distribution.

Calculating the emissivities and absorptivities thus amounts to carrying out the integration over momentum space and sum over harmonics $n$ in equations \eqref{eq:EmiStokes} and \eqref{eq:AbsStokes}.  

\section{Electron Distribution Functions} \label{sec:DF}

The synchrotron emissivity and absorptivity depends on the electron distribution function:
\begin{equation}
f \equiv \frac{dn_e}{d^3p} = \frac{1}{m_e^3 c^3 \gamma^2 \beta} \frac{dn_e}{d\gamma d\cos\xi d\phi}.
\end{equation}
Here $n_e$ is the electron number density, and we use the electron momentum space coordinates $\gamma, \xi, \phi$; as usual $\gamma$ is the electron Lorentz factor, $\xi$ is the pitch angle, and $\phi$ is the gyrophase.

The numerical scheme described below can calculate emissivities and absorptivities for an arbitrary electron distribution function that is independent of gyrophase.  We investigate three commonly used isotropic DFs: a relativistic thermal (Maxwell-J\"uttner) distribution; a power-law distribution; and a kappa distribution. 

\subsection{Thermal distribution}

The {\em thermal} distribution function is:
\begin{equation}
\frac{dn_e}{d\gamma d\cos\xi d\phi} = \frac{n_e}{4 \pi \Theta_e} \frac{\gamma (\gamma^2 - 1)^{1/2}}{K_2(1/\Theta_e)} \text{ exp}\left(- \frac{\gamma}{\Theta_e}\right),
\end{equation}
where $\phi$ is the gyrophase, $K_2$ is a modified Bessel function of the second kind, and $\Theta_e \equiv k_B T/m_e c^{2}$ is the dimensionless temperature.  The nonrelativistic limit ($\Theta_e \ll 1$) is 
\begin{equation}
\frac{dn_e}{d^3 {\rm v}} = n_e \left(\frac{m_e}{2 \pi k T_e}\right)^{3/2} e^{-v^2 m_e/(2 k T_e)}.
\end{equation}
where $d\gamma d\cos\xi d\phi \approx d^3{\rm v} /(c^2 v)$, and $K_2(\Theta_e^{-1}) \approx  (\pi \Theta_e/2)^{1/2} \exp(-1/\Theta_e)$ for $\Theta_e \ll 1$.

The thermal distribution is widely used \citep[e.g.][]{Pacholzyk1970, Takahara1982, Mahadevan1996, Mosc2009, Leung2011}. It is unlikely, however, that the distribution function is able to thermalize in the collisionless conditions typical of many synchrotron-emitting plasmas, so it is useful to have a nonthermal model as well.

\subsection{Power-law distribution}

A commonly used nonthermal model is the {\em power-law} distribution function:
\begin{equation}
\frac{dn_e}{d\gamma d\cos{\xi} d\phi} = \frac{n^{NT}_e (p-1)}{4 \pi (\gamma^{1-p}_{min} - \gamma^{1-p}_{max})} \gamma^{-p} \qquad \text{        for } \gamma_{min} \leq \gamma \leq \gamma_{max},
\end{equation}
and zero otherwise.  Here $n^{NT}_e$ is the number density of nonthermal electrons, and $\gamma_{min}$, $\gamma_{max}$, and p are parameters of the distribution.  

The power-law distribution is well motivated in the sense that observations directly imply the existence of power-law tails in distant synchrotron-emitting plasmas, in the solar wind, and in numerical particle-in-cell experiments studying particle acceleration in reconnection and shocks.  Nevertheless if $\gamma_{min} = 1$ then most of the electrons are clustered near $\gamma = 1$; if $\gamma_{min} \ne 1$ then the distribution function has a central hole that makes it unstable (to the bump-on-tail instability; see \cite{Kulsrud2005}).  It is therefore useful to have a compromise distribution available that is stable, has a thermal core, and asymptotes to a power-law at high energy. This motivates the {\em kappa} distribution function.

\subsection{Kappa distribution}

\begin{figure}
    \centering
    \includegraphics[width=0.9\textwidth]{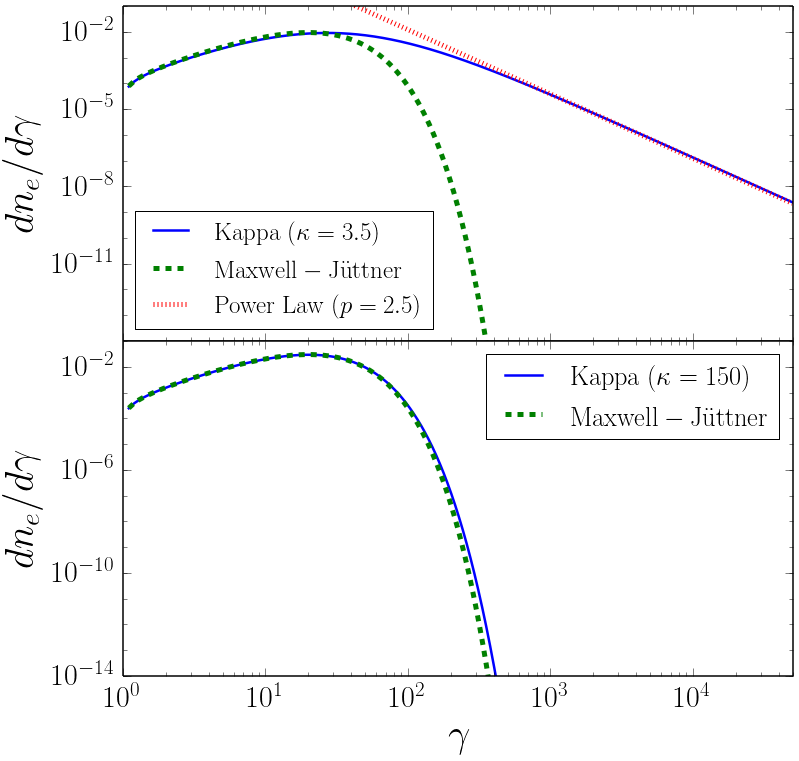}
    \caption{Plot of the kappa distribution for $\kappa = 3.5$ and width $w = 10$. The $\kappa$ distribution asymptotes to the thermal distribution, with $\Theta_e = w = 10$, at low particle Lorentz factor $\gamma$ and the power law distribution, with $p = \kappa -1 = 2.5$ at large $\gamma$, and smoothly connects the two.}
    \label{fig:kappavsthemalinsection3}
\end{figure}

The kappa distribution consists of a nonthermal, power-law tail at large $\gamma$ that smoothly transitions to a flat, thermal-like core for small $\gamma$. It originated as a fit to observed solar wind data (\citealt{Vasyliunas1968}, see \citealt{Pierrard2010} for a recent review). One interesting property of the kappa distribution is that it corresponds to the maximum of a modified entropy functional, namely the Tsallis non-extensive entropy (\citealt{Tsallis1988}). The functional form of the distribution can be obtained by extremizing this entropy functional, subject to thermodynamic constraints, such as density and an appropriate definition of temperature (\citealt{Livadiotis2009}). While this method gives the functional form of the distribution, $\kappa$ is still an input parameter. In the limit $\kappa \rightarrow \infty$, the kappa distribution asymptotes to the thermal distribution. Figure (\ref{fig:kappavsthemalinsection3}a) shows how the kappa distribution function connects the thermal and the power law distribution functions, and figure (\ref{fig:kappavsthemalinsection3}b) shows the $\kappa \rightarrow \infty$ limit of the kappa distribution along with a thermal distribution.

The relativistic {\em kappa} distribution function (\citealt{Xiao2006}) is :
\begin{equation} \label{eq:rel_kappa}
\frac{dn_e}{d\gamma d\cos{\xi} d\phi} = \frac{N}{4 \pi} \gamma (\gamma^2 - 1)^{1/2} \left(1 + \frac{\gamma-1}{\kappa w}\right)^{-(\kappa+1)},
\end{equation}
where $n_e$ is the number density of electrons, and $\kappa$ and $w$ are parameters.  $N$ is the normalization, which can be evaluated analytically but involves special functions and is sufficiently  complicated as not to be very useful.  In certain limits $N$ simplifies to
\begin{equation}
    N(\kappa, w)= 
\begin{cases}
    n_{e}(\frac{2}{\pi \kappa^3 w^3})^{1/2} \frac{\Gamma(\kappa+1)}{\Gamma(\kappa-1/2)},& \text{if } \kappa w \ll 1 \\
    n_{e} \frac{(\kappa-2)(\kappa-1)}{2 \kappa^2 w^3},&\text{if } \kappa w \gg 1,
\end{cases}
\end{equation}

Typically we evaluate $N$ numerically.\footnote{The small $\kappa w$ limit $N_-$ and the high $\kappa w$ limit $N_+$ can be combined to yield $N \approx (N_-^{-0.7} + N_+^{-0.7})^{1/0.7}$ which has maximum error for $w \simeq 1/2$ of $((4, 1.8, 0.8))\%$ at $\kappa = ((3, 4, 10))$.}  

In the nonrelativistic limit $w\kappa \ll 1$, (\ref{eq:rel_kappa}) asymptotes to:
\begin{equation}
\frac{dn_e}{d^3{\rm v}} = \frac{n_e}{(\pi \kappa w_{NR}^2)^{3/2}} \, \frac{\Gamma(\kappa+1)}{\Gamma(\kappa - 1/2)} \, \left(1 + \frac{v^2}{\kappa w_{NR}^2}\right)^{-(\kappa+1)}
\end{equation}
where $d\gamma d\cos\xi d\phi \approx d^3{\rm v} /(c^2 v)$, $w_{NR}^2 \equiv 2 c^2 w$.

The parameter $w$ is analogous to $\Theta_e$, and is a measure of dispersion in momentum space. In the limit $\kappa \rightarrow \infty$, $(1 + (\gamma - 1)/(\kappa w))^{-(\kappa+1)} \rightarrow \exp(-(\gamma-1)/w)$ in (\ref{eq:rel_kappa}) and hence the kappa distribution goes over to the thermal distribution, with $w = \Theta_e$ (normalization takes care of the extra factor of $\exp(-1/w)$). Typically, $\kappa$ is of order unity, so $\kappa w \ll 1$ is the nonrelativistic limit and $\kappa w \gg 1$ is the ultrarelativistic limit.

\section{Numerical Scheme}\label{sec:methods}

\subsection{Formulation}
\begin{figure}
    \centering
    \includegraphics[width=0.9\textwidth]{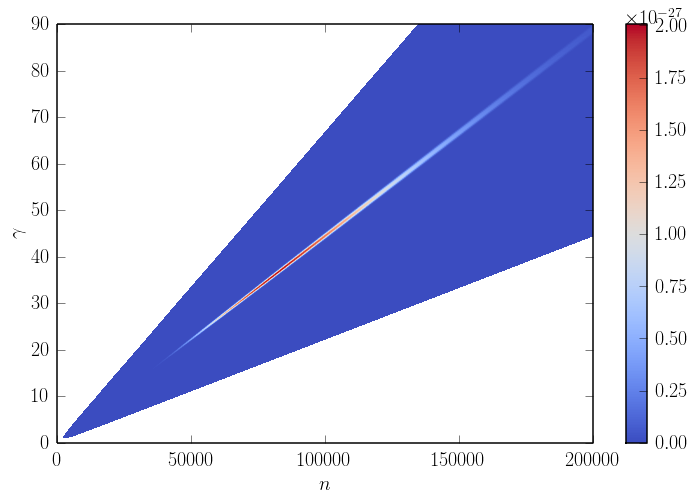}
    \caption{The integrand $I(n,\gamma)$ in the $n$-$\gamma$ plane for the synchrotron emissivity of a thermal distribution with $\Theta_e = 10$, $\theta = 60$deg, B = 30 G and $\nu/\nu_c = 3000$. The location of the peak corresponds to the line $\gamma = n/((\nu/\nu_c) \sin^2\theta)$ (see \S \ref{sec:algorithm} for derivation).}
    \label{fig:integrandThermalAbs}
\end{figure}

The general expressions for both the emissivities \eqref{eq:EmiStokes} and the absorptivities \eqref{eq:AbsStokes} for a gryotropic distribution function reduce to a set of integrals and a summation of the form:
\begin{equation}
\int_1^\infty d\gamma \int_{-1}^1 d\cos\xi \, \sum_{n=1}^\infty\,\, \delta(y_n) \, I(n,\xi,\gamma).
\end{equation}
Here $I$ is the integrand (not the Stokes parameter!).  The $\delta$ function can be eliminated by integrating over one of $\gamma, \xi$, or $n$.  As in \cite{Leung2011} we integrate over $\cos\xi$.  Setting $y_n = 0$ implies the substitution
\begin{equation}\label{eq:deltasub}
\cos\xi = \frac{\nu - n \nu_c/\gamma}{\nu\beta\cos\theta}
\end{equation}
The ranges of integration and summation are then restricted by the conditions that $1 \le \gamma < \infty$ and $-1 \le \cos\xi \le 1$.  The final form of the integrand is 
\begin{equation}\label{eq:igrand}
\int_{\gamma_-}^{\gamma_+} d\gamma \sum_{n=n_-}^\infty\,\, I(n,\gamma)
\end{equation}
where $\gamma_\pm = (r \pm |\cos\theta|(r^2 - \sin^2\theta)^{1/2})/\sin^2\theta$, $r = n\nu_c/\nu$,  and $n_- = (\nu |\sin\theta| /\nu_c)$.  Figure (\ref{fig:integrandThermalAbs}) shows a heat map of the integrand $I(n,\gamma)$ for the emissivity of a thermal distribution.

The integrands are either reflection symmetric or antisymmetric about $\theta = 90$deg.  It is possible to show from equations \eqref{eq:KIdef}-\eqref{eq:zdef}, 
\eqref{eq:deltasub}, \eqref{eq:igrand} that $K_I$ is symmetric, $K_Q$ is symmetric, and $K_V$ is antisymmetric.  This implies that the sense of emitted circular polarization changes moving from more nearly parallel to more nearly antiparallel to the field, and that the absorption coefficient has the same symmetry.  

\subsection{Double Integration Algorithm}\label{sec:algorithm}

We shall now describe the scheme that computes \eqref{eq:igrand}.  We implement the scheme in a new code, {\tt symphony}.  It differs from the {\tt harmony} code of Leung et al. in code organization (our code is vastly simplified), as well as in technical aspects of the integration.  

As in Leung et al., the summation is done directly for $n < 30$, and is approximated by an integral for $n \ge 30$.  The $\gamma$ integral is evaluated first, followed by the $n$ summation/integration. The integrals are evaluated using the GNU Science Library (GSL) Quasi-Adaptive Gaussian quadrature routines QAG and QAGIU, with accuracy controlled by a relative error tolerance of $10^{-3}$ and $10^{-8}$, respectively.  

The $n$ integral is performed well by the method described by \cite{Leung2011}, but the $\gamma$ integration method breaks down at large $\nu/\nu_c$, where the integrand is sharply peaked enough that QAG can miss the peak entirely and return $0$. This problem is solved by adaptively narrowing the range of integration based on an estimate for the location and width of the peak in $\gamma$-space. The location of the peak corresponds (using $y_n = 0$, $\gamma \gg 1$, and assuming that $\theta$ is not close to $0$ or $\pi$) to the line $\gamma = n/((\nu/\nu_c) \sin^2\theta)$.  This estimate is based on the physical notion that in the ultrarelativistic limit emission at observer angle $\theta$ originates from electrons in a narrow cone of pitch angles of width $1/\gamma$ around $\theta \approx \xi$.  Notice that the emissivity and absorptivity are in any case susceptible to accurate asymptotic approximation in this limit.

The Leung et al. method also fails for Stokes V because the $\gamma$ integrand looks similar to a period of the sine function, with one lobe slightly larger than the other; QAG has difficulty resolving this small difference in area. To avoid erroneous numerical cancellation in this integral, we integrate from the leftmost bound to the zero in the middle of the sinusoid-like curve, and then sum/integrate this piece over $n$.  We then follow the same procedure for the right side of the sinusoid-like curve, and then sum the two results to get the final emission or absorption coefficient. This is not difficult because the integrand goes through one and only one zero at $\cos\theta = \beta \cos\xi$, which corresponds to a zero at $\gamma_0 = n \nu_c/(\nu \sin^2 \theta)$.  When $\gamma \gg 1$ this corresponds to $\theta \simeq \xi$, and this has the simple physical interpretation that the sign of $K_V$ (and hence $j_V$ and $\alpha_V$) depends on the sense of rotation of the electron (orbit described by $\xi$) around the wavevector (direction described by $\theta$).

\subsection{Tests}

We have verified {\tt symphony} by comparing to {\tt harmony}, to existing fitting formulae, and to asymptotic expressions for the emissivity and absorptivity. In all of the numerical comparisons that follow we assume, unless otherwise stated, that $B = 30$ Gauss, $\Theta_e = 10$, $\theta = 60$deg, and $\nu/\nu_c = 10^2$.

\subsection{Thermal distribution: Emissivity and Absorptivity}

\cite{Leung2011} introduced a fitting formula for the relativistic thermal synchrotron emissivity:
\begin{equation}\label{eq:thermJIfit}
j_{I}=n_{e}\frac{\sqrt{2}\pi e^{2}\nu_{s}}{3K_{2}(1/\Theta_{2})c}(X^{1/2}+2^{11/12}X^{1/6})^{2}\exp(-X^{1/3}),
\end{equation}
where $X \equiv \nu/\nu_{s}$ and $\nu_{s}\equiv (2/9)\nu_{c}\Theta_{e}^{2}\sin{\theta}$. 

\begin{figure}
    \centering
    \includegraphics[width=0.9\textwidth]{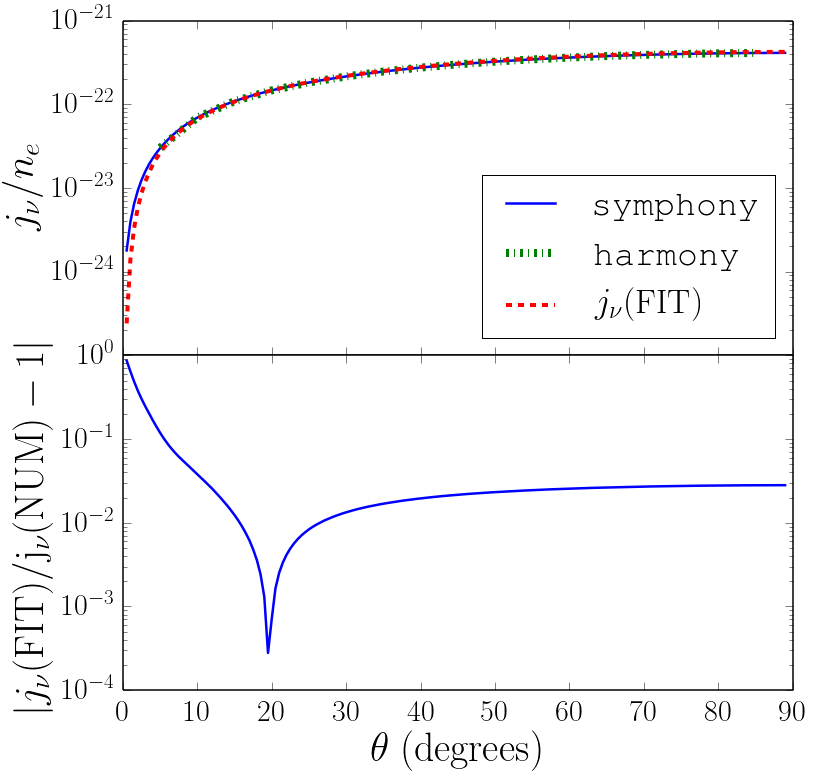}
    \caption{Comparison of {\tt symphony}, {\tt harmony}, and the fitting formula for Stokes I thermal synchrotron emission (equation \eqref{eq:thermJIfit}) versus observer angle $\theta$.  The parameters used are $\Theta_e = 10$, $B = 30$G, and $\nu/\nu_c = 10^2$.}
\label{fig:thermaltestemiangle}
\end{figure}

\begin{figure}
    \centering
    \includegraphics[width=0.9\textwidth]{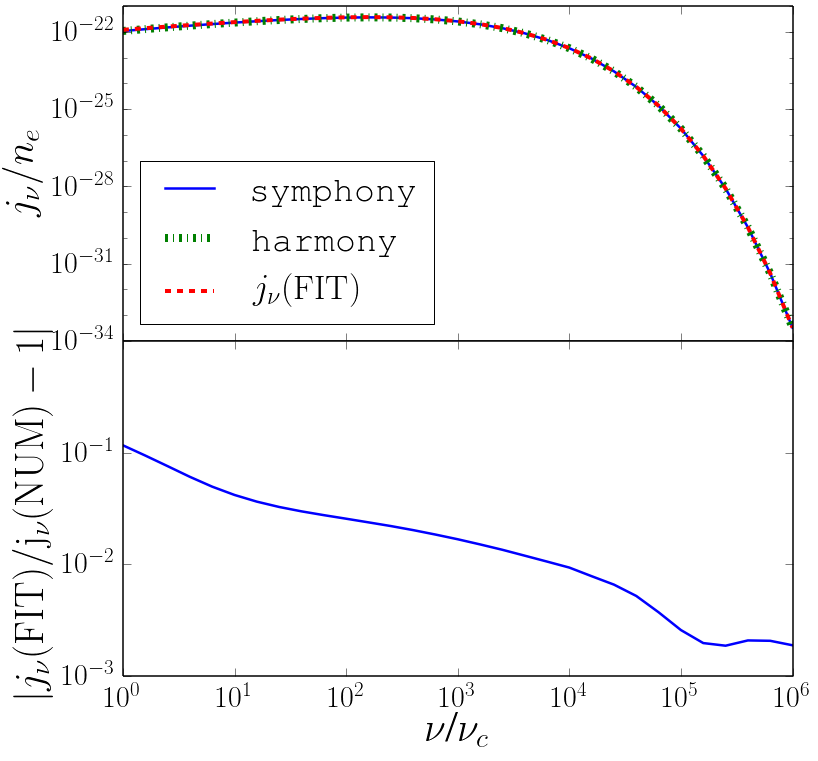}
    \caption{Comparison of {\tt symphony}, {\tt harmony}, and the fitting formula for Stokes I thermal synchrotron emission (equation \eqref{eq:thermJIfit}) versus frequency in terms of $\nu/\nu_c$. The parameters used are $\Theta_e = 10$, $B = 30$G, and $\theta = 60$deg.}
    \label{fig:thermaltestemifreq}
\end{figure}

We tested {\tt symphony} against both the fitting formula and {\tt harmony}.  We find good agreement over the range shown in \cite{Leung2011}; a partial comparison is shown in figures (\ref{fig:thermaltestemiangle}) and (\ref{fig:thermaltestemifreq}).

The thermal distribution is special in that the emission and absorption coefficients must be related by Kirchoff's law.  In the Stokes basis Kirchoff's law reads
\begin{equation}\label{eq:kirchoff}
J_{S} - \alpha_{S} B_{\nu} = 0,
\end{equation}
where $B_{\nu} \equiv (2h\nu^{3}/c^{2})[\exp(h\nu/kT_{e}) - 1]^{-1}$ is the Planck function.  Equations \eqref{eq:thermJIfit} and \eqref{eq:kirchoff} imply a fitting formula for the thermal absorptivity $\alpha_I$.  We tested our code against this formula and against {\tt harmony} and once again find good agreement, with a maximum fractional error $10^{-3}$ for $\nu/\nu_c = 1$.

\subsection{Power Law distribution: Emissivity, Absorptivity and Polarization}
For the power-law distribution
\begin{equation}\label{eq:plJIemiss}
j_{\nu}=n_{e}^{NT}\left(\frac{e^{2}\nu_{c}}{c}\right)\frac{3^{p/2}(p-1)\sin{\theta}}{2(p+1)(\gamma_{min}^{1-p}-\gamma_{max}^{1-p})}\Gamma\left(\frac{3p-1}{12}\right)\Gamma\left(\frac{3p+19}{12}\right)\left(\frac{\nu}{\nu_{c}\sin{\theta}}\right)^{-(p-1)/2}
\end{equation}
and the absorptivity is \citep[e.g.][]{Rybicki1979}
\begin{equation}\label{eq:plJIabs}
\alpha_{\nu}=n_{e}^{NT}\left(\frac{e^{2}}{\nu m_{e}c}\right)\frac{3^{p/2}(p-1)}{4(\gamma_{min}^{1-p}-\gamma_{max}^{1-p})}\Gamma\left(\frac{3p+12}{12}\right)\Gamma\left(\frac{3p+22}{12}\right)\left(\frac{\nu}{\nu_{c}\sin{\theta}}\right)^{-(p+2)/2}.
\end{equation}
Both these expressions are obtained in the ultrarelativistic limit and are strictly valid only for $\nu \gg \nu_c$.  

\begin{figure}
    \centering
    \includegraphics[width=0.9\textwidth]{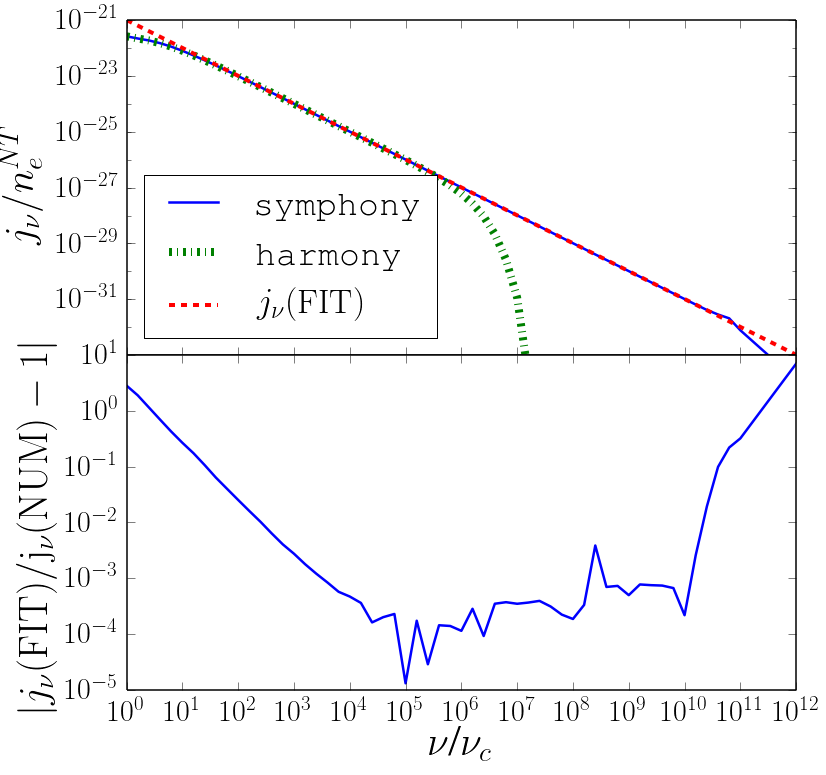}
    \caption{Failure of {\tt harmony} at high $\nu/\nu_c$ for a power-law distribution. Here $p = 3$, $\theta = 60$deg and $B = 100$.}
    \label{fig:powerlawtestemi}
\end{figure}

We tested {\tt symphony} against {\tt harmony} and \eqref{eq:plJIemiss} and \eqref{eq:plJIabs}. {\tt harmony} matches the fitting formula only up to $\nu / \nu_c = 10^6$, whereas {\tt symphony} gives the right result throughout $10^1 < \nu/\nu_c < 10^{10}$. This is because, as mentioned earlier, the integration strategy of Leung et al. fails to capture the narrow peak of the integrand in $\gamma$ in this regime.  {\tt symphony} adopts a modified strategy that narrows the range of integration, but even this begins to fail at $\nu/\nu_c > 10^{10}$. This is shown in figure (\ref{fig:powerlawtestemi}).

\begin{figure}
    \centering
    \includegraphics[width=0.9\textwidth]{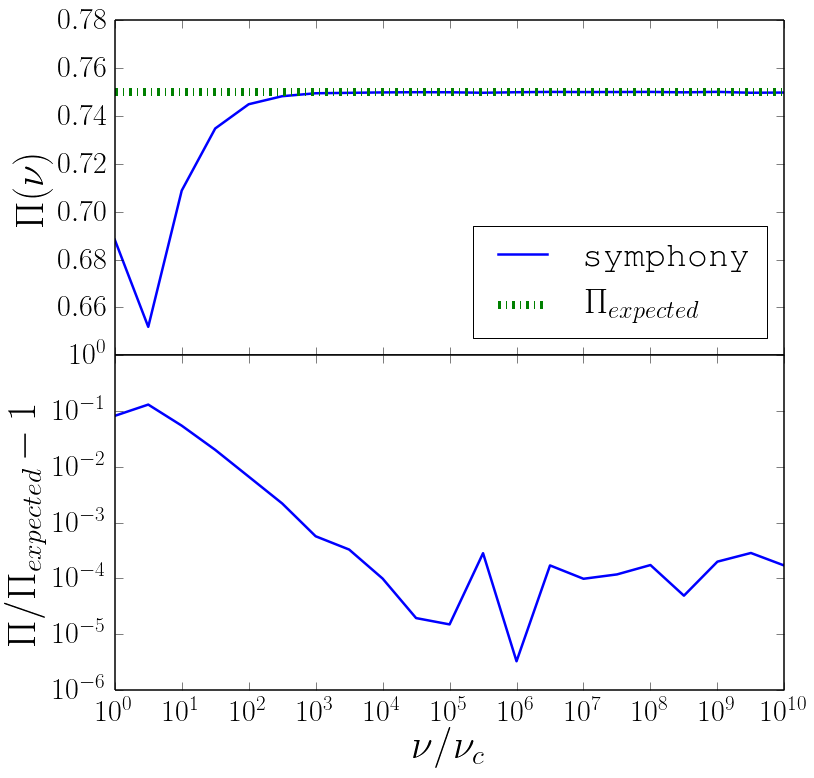}
    \caption{Polarization fraction calculated in {\tt symphony} for a power-law distribution with $p = 3$.  The expected fraction is $0.75$, according to the asymptotic analysis of \cite{Legg1968}.}
    \label{fig:polar3}
\end{figure}

The power-law distribution is known to produce linear polarization that depends on the power-law index:
\begin{equation}\label{eq:plpolfrac}
\Pi(\nu) = \frac{|J_Q(\nu)|}{J_I(\nu)}= \frac{p+1}{p+\frac{7}{3}}
\end{equation}
\citep{Legg1968}.
Figure (\ref{fig:polar3}) shows that our numerical results agree with this expression for sufficiently large $\nu/\nu_c$ with a difference that is entirely from truncation error.  At low $\nu/\nu_c$ the analytic estimate is higher than {\tt symphony} by a few percent.  This difference is physical and shows the limits of the asymptotic approximation used in deriving the polarization fraction.  Calculations for other $p$ yield similar results.

To sum up: there is good agreement between our new method and our earlier, extensively tested code, and also between our method and asymptotic expressions for the emissivity, absorptivity, and polarization fraction.  

\section{Results} \label{sec:results}

\subsection{Emissivites and Absorptivities}

\begin{figure}
    \centering
    \includegraphics[width=0.495\textwidth]{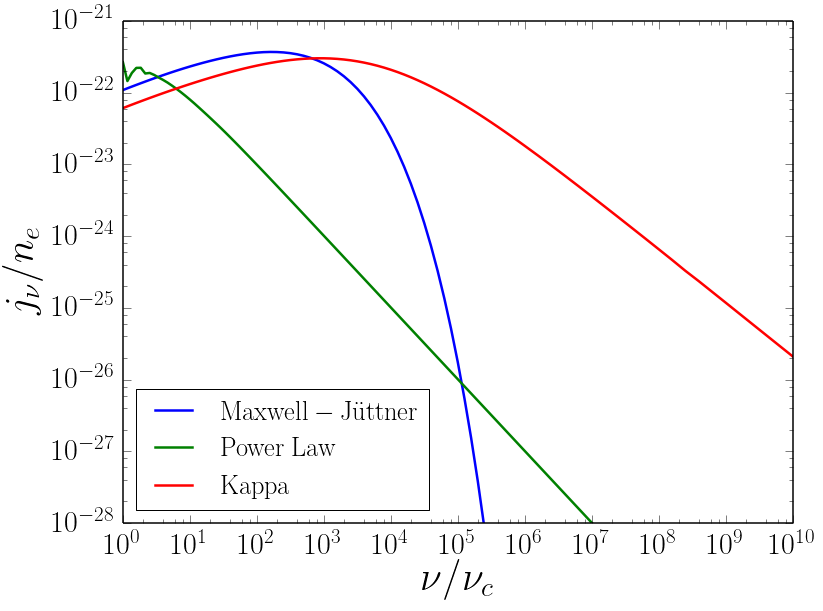}
    \includegraphics[width=0.495\textwidth]{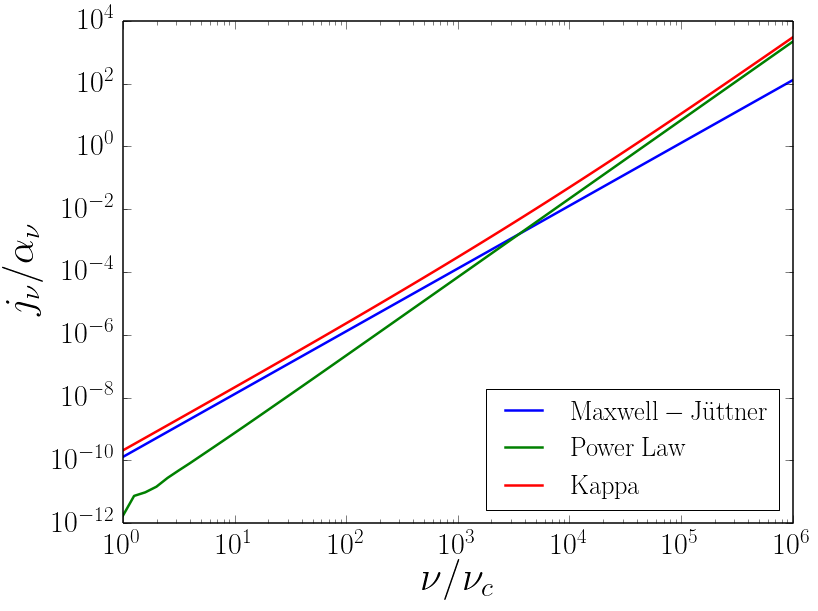}
    \caption{(left) Emissivity in Stokes I for thermal ($\Theta_e = 10$), power-law ($p=3$) and kappa distributions ($\kappa = 3.5, w = 10$) as a function of frequency.  The cyclotron line structure at $\nu/\nu_c \sim 1$ in the power-law distribution is real and a consequence of the large number of mildly relativistic electrons in a power-law distribution with $\gamma_{min} = 1$.  (right) Source function ($j_I/\alpha_I$) for power-law, thermal, and kappa distributions as a function of frequency.}
    \label{fig:allspectra}
\end{figure}

Having tested {\tt symphony} for a wide range of parameters for both the thermal and the power law distribution, we can now use it to probe whether our three electron distribution functions produce significantly different spectra.  Figure (\ref{fig:allspectra}) compares representative emissivities $j_I$ and source functions ($j_I/\alpha_I$) as a function of frequency, with all other parameters set to their standard values. 

Evidently the kappa distribution looks like the thermal distribution at low frequency and a power-law distribution at high frequency.  Notice that the source function for the kappa distribution is indistinguishable from thermal at low frequency, and only regains the characteristic nonthermal self-absorbed form, $\propto \nu^{5/2}$, above a characteristic frequency $\nu_\kappa \sim (w\kappa)^2 \nu_c$.  

\subsection{Polarization}

\begin{figure}
    \centering
    \includegraphics[width=0.475\textwidth]{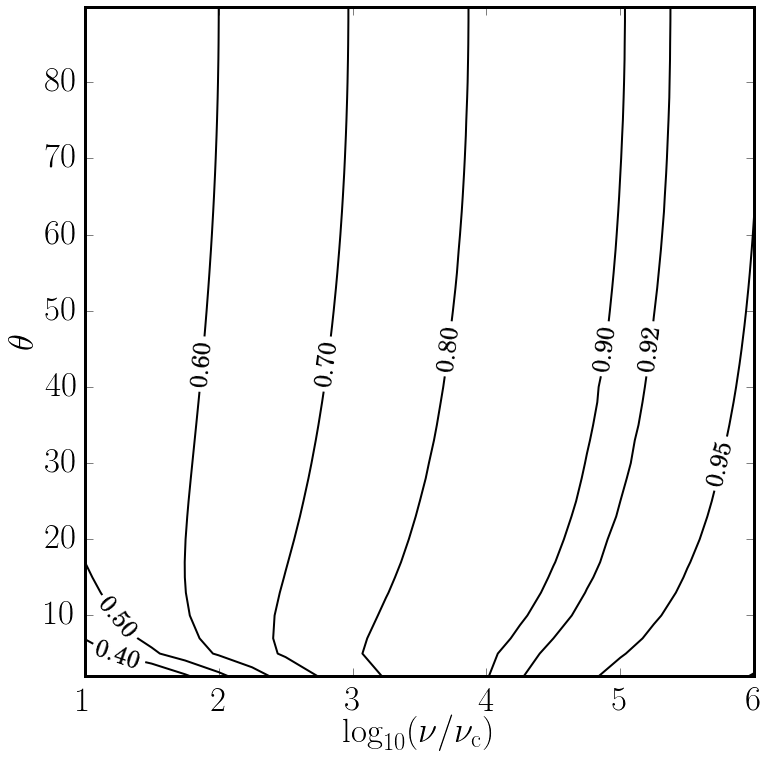}
    \includegraphics[width=0.475\textwidth]{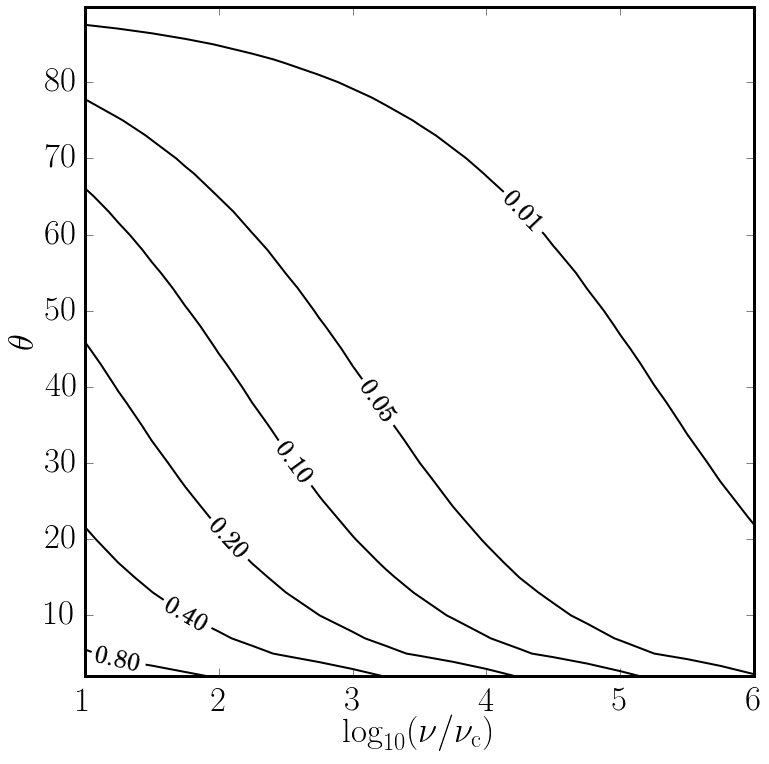}
    \caption{Polarization fraction for a thermal distribution function.  (left) Fractional linear polarization $|j_Q|/j_I$.  (right) Fractional circular polarization $|j_V|/j_I$.  Here and below, only $0 \le \theta \le 90$deg is shown; $90 \le \theta \le 180$deg can be obtained by symmetry.}
    \label{fig:thermalpol}
\end{figure}

\begin{figure}
    \centering
    \includegraphics[width=0.475\textwidth]{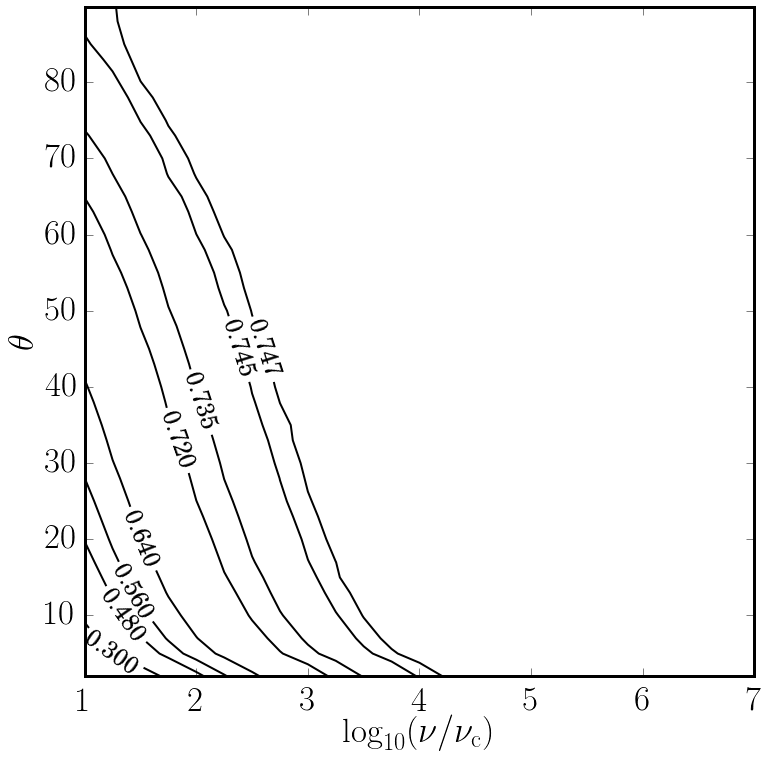}
    \includegraphics[width=0.475\textwidth]{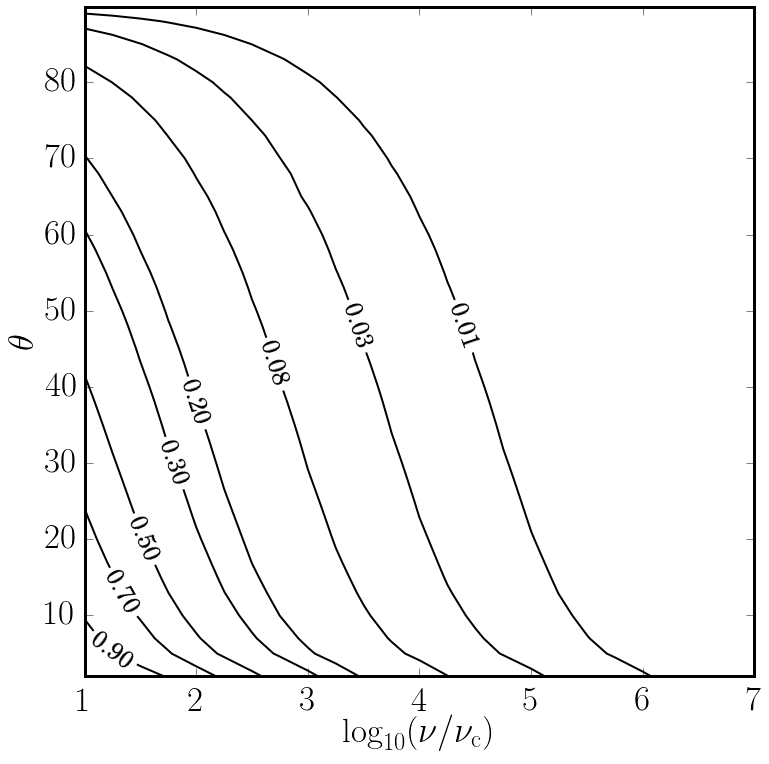}
    \caption{Polarization fraction for a power-law distribution function.  (left) Fractional linear polarization $|j_Q|/j_I$.  (right) Fractional circular polarization $|j_V|/j_I$.}
    \label{fig:powerpol}
\end{figure}

\begin{figure}
    \centering
    \includegraphics[width=0.475\textwidth]{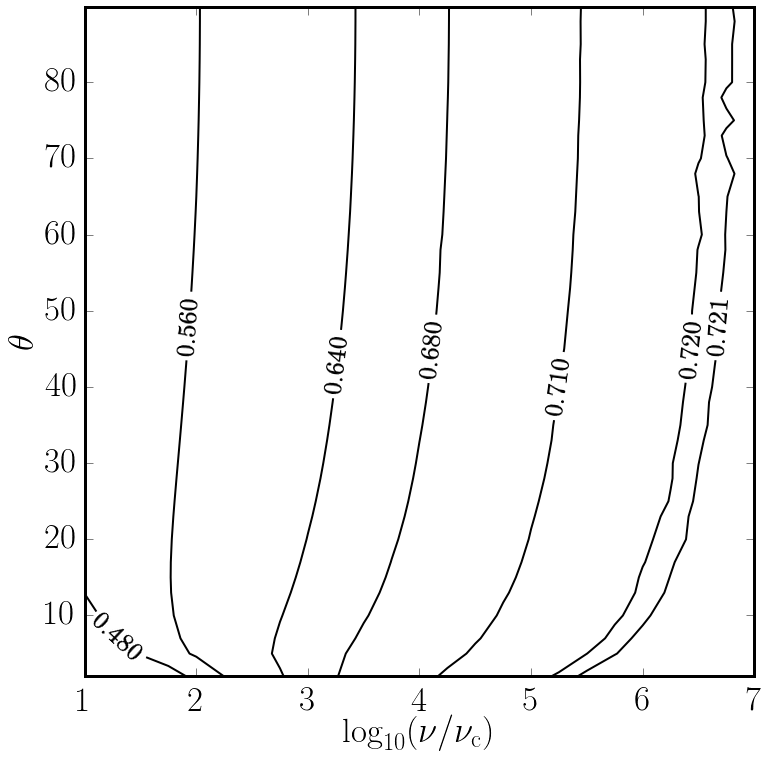}
    \includegraphics[width=0.475\textwidth]{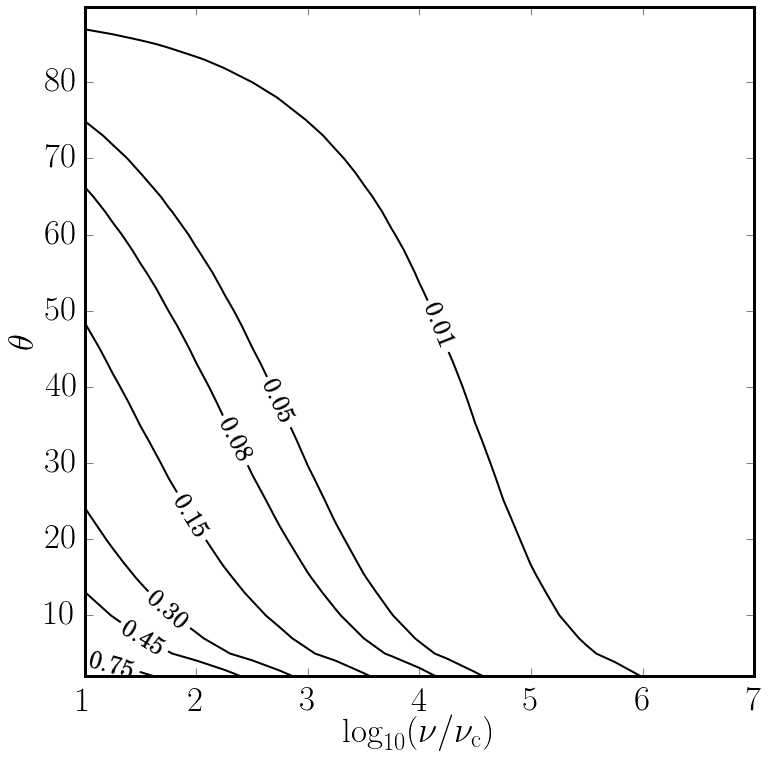}
    \caption{Polarization fraction for a kappa  distribution function.  (left) Fractional linear polarization $|j_Q|/j_I$.  (right) Fractional circular polarization $|j_V|/j_I$.}
    \label{fig:kappapol}
\end{figure}

Are there significant differences in the polarization properties of emitted radiation from the different distributions?  Figures (\ref{fig:thermalpol}), (\ref{fig:powerpol}), and (\ref{fig:kappapol}) show contour plots of the emitted linear and circular polarization fraction as a function of $\nu/\nu_c$ and $\theta$ for all three distribution functions, with our usual representative parameters.  

In the figures we only show polarization fractions for $\theta \le 90$deg; the polarization at $\theta \ge 90$deg can be obtained from this by symmetry arguments.  For Stokes I it is easy to show that the two hemispheres are symmetric, and likewise for Stokes Q.  On the other hand, Stokes V is antisymmetric about $\theta = 90$deg (this can be proved directly from the emissivity formula, once the $\delta$ function is taken into account).  

Notice that all distribution functions have reduced linear polarization at small $\nu/\nu_c$ (for the parameters chosen here, less than about $10^2$).  In addition, all distribution functions have some expected circular polarization at low $\nu/\nu_c$ and away from the $\theta = 90$deg plane.  The highest circular polarization is obtained on lines of sight nearly aligned with the field ($\theta = 0$), but this is also where the emission is lowest.  

Perhaps most surprising is the high degree of linear polarization of the thermal distribution, which approaches 100\% at high frequency.  The emission comes from a narrow band of Lorentz factors with $\gamma \sim (\Theta_e \nu/\nu_c)^{1/3}$ when $\nu \gg \Theta_e^2 \nu_c$.  This happens because even though these lower energy electrons radiate weakly at high energy, there are exponentially more of them than there are electrons at the usual $\gamma \sim (\nu/\nu_c)^{1/2}$.  It is easy to show that the linear polarization fraction associated with monoenergetic electrons approaches $1$ as $\nu/(\gamma^2 \nu_c) \rightarrow \infty$, and therefore that the polarization fraction of the thermal distribution should go to $1$ for $\nu/(\Theta_e^2 \nu_c) \gg 1$.  The high degree of polarization is found not just in the thermal distribution but at high frequency in any distribution function with an exponential cutoff.

\subsection{Fitting Formulae}

For some applications, the numerical integration is too computationally expensive. In this case it is useful to generate a lookup table, or to derive fitting formulae.  Below we provide fitting formulae for absorptivity and emissivity in all the Stokes parameters, for the three distribution functions discussed so far.  These fitting formulae are implemented in the public code release, alongside the numerical integration scheme.

Synchrotron emissivities in vacuum have the universal form 
\begin{equation}
j_{S} = \frac{n_e e^2 \nu_c}{c} J_S\left(\frac{\nu}{\nu_c}, \theta\right),
\end{equation}
and absorptivities have the universal form
\begin{equation}
\alpha_{S} = \frac{n_e e^2}{\nu m_e c} A_S\left(\frac{\nu}{\nu_c}, \theta\right)
\end{equation}
where $J_S$ and $A_S$ are dimensionless and depend on the distribution function parameters: $\Theta_e$ (thermal), $p$ (power-law, assuming the upper and lower cutoffs are absent), and $w,\kappa$ (kappa).  It is easily confirmed that $\alpha_S$ and $j_S$ have the correct dimensions.

\subsubsection{Thermal Distribution}

The dimensionless emissivity is
\begin{equation}
J_S = e^{-X^{1/3}} 
\begin{cases}
\frac{\sqrt{2} \pi}{27} \sin(\theta)(X^{1/2}+2^{11/12}X^{1/6})^{2}, & \text{Stokes I} \\
-\frac{\sqrt{2} \pi}{27} \sin(\theta)(X^{1/2}+ \left(\frac{7 \Theta_{e}^{24/25} + 35}{10 \Theta_{e}^{24/25} + 75}  \right) 2^{11/12}X^{1/6})^{2}, & \text{Stokes Q} \\
0, & \text{Stokes U} \\
-\frac{37-87\sin(\theta-\frac{28}{25})}{100(\Theta_e + 1)} \left(1 + \left(\frac{\Theta_{e}^{3/5}}{25} + \frac{7}{10} \right) X^{9/25} \right)^{5/3}, & \text{Stokes V}
\end{cases}
\end{equation}
The dimensionless absorptivity is 
\begin{equation}
A_S = \frac{J_S}{B_\nu} = J_S \,\, \frac{m_e c^2 \nu_c}{2 h \nu^2} \,\,
\left(e^{h\nu/(k T)} - 1\right)
\end{equation}
using Kirchoff's law.

\subsubsection{Power-Law Distribution}

The dimensionless emissivity is 
\begin{multline}
J_S = 
\frac{3^{p/2}(p-1)\sin{\theta}}{2(p+1)(\gamma_{min}^{1-p}-\gamma_{max}^{1-p})}\Gamma\left(\frac{3p-1}{12}\right)\Gamma\left(\frac{3p+19}{12}\right)\left(\frac{\nu}{\nu_{c}\sin{\theta}}\right)^{-(p-1)/2}
\\ \, \, \times \, \,
\begin{cases}
1, & \text{Stokes I} \\
-\frac{p+1}{p+7/3}, & \text{Stokes Q} \\
0, & \text{Stokes U} \\
-\frac{171}{250} \frac{p^{1/2}}{\tan(\theta)} (\frac{\nu}{3 \nu_c \sin(\theta)})^{-1/2}, & \text{Stokes V}
\end{cases} \qquad
\end{multline}

The dimensionless absorptivity cannot be obtained from Kirchoff's law, and must be fit separately:
\begin{multline}
A_S =\frac{3^{(p+1)/2}(p-1)}{4(\gamma_{min}^{1-p}-\gamma_{max}^{1-p})}\Gamma\left(\frac{3p+12}{12}\right)\Gamma\left(\frac{3p+22}{12}\right)\left(\frac{\nu}{\nu_{c}\sin{\theta}}\right)^{-(p+2)/2} \\
\times \, \,
\begin{cases}
1, & \text{Stokes I} \\
-\frac{3}{4}(p - 1)^{43/500}, & \text{Stokes Q} \\
0, & \text{Stokes U} \\
-\frac{7}{4}(\frac{71}{100}p + \frac{22}{625})^{197/500}((\sin{\theta})^{-48/25} - 1)^{64/125} (\frac{\nu}{\nu_c\sin{\theta}})^{-1/2}, & \text{Stokes V}
\end{cases} \qquad
\end{multline}
These fits are suitable for $\gamma_{min}^2 < \nu/\nu_c < \gamma_{max}^2$.

\subsubsection{Kappa Distribution}

The kappa distribution is characterized by an approximately thermal core and power-law tails with power-law parameter $p = \kappa - 1$.  Figure (\ref{fig:kappavsthemalinsection3}) shows kappa distributions with various values of $\kappa$, as well as the Maxwell-J\"uttner distribution.

Our fitting formulae are more complicated for the kappa distribution function than for the power-law and thermal DFs.  We proceed by identifying a low-frequency and high-frequency fit, usually inspired by asymptotic expansions, and then introduce bridging formulae that interpolates between them. In what follows it is useful to define the characteristic frequency $\nu_\kappa \equiv \nu_c (w\kappa)^2 \sin\theta$, and $X_\kappa \equiv \nu/\nu_\kappa$.

In the low frequency limit
\begin{multline}
J_{S,lo} =  X^{1/3}_\kappa
\sin(\theta)\frac{4\pi\Gamma(\kappa-4/3)}{3^{7/3}\Gamma(\kappa-2)}  \\
\, \times \, 
\begin{cases}
1 & \text{Stokes I} \\
-\frac{1}{2} & \text{Stokes Q} \\
0 & \text{Stokes U} \\
-(\frac{3}{4})^{2}\left[(\sin{\theta})^{-12/5} - 1\right]^{12/25}\frac{\kappa^{-66/125}}{w} X_{\kappa}^{-7/20} & \text{Stokes V}
\end{cases} \qquad
\end{multline}
While in the high frequency limit
\begin{multline}
J_{S,hi} =  X^{-(\kappa-2)/2}_\kappa
\sin(\theta)\, 3^{(\kappa-1)/2} \frac{(\kappa-2)(\kappa-1)}{4} \Gamma(\frac{\kappa}{4}-\frac{1}{3}) \Gamma(\frac{\kappa}{4} + \frac{4}{3}) \\
\, \times \,
\begin{cases}
1, & \text{Stokes I} \\
(-1)[(\frac{4}{5})^{2} + \frac{1}{50}\kappa] & \text{Stokes Q} \\
0 &\text{Stokes U} \\
-(\frac{7}{8})^{2}\left[(\sin{\theta})^{-5/2} - 1\right]^{11/25} \frac{\kappa^{-11/25}}{w} X_{\kappa}^{-1/2} & \text{Stokes V}
\end{cases} \qquad
\end{multline}

Then the bridging function is :
\begin{equation}
J_S = \left(J_{S,lo}^{-x}+ J_{S,hi}^{-x}\right)^{-1/x},
\end{equation}
where
\begin{equation}
x = 
\begin{cases}
3\kappa^{-3/2}, &\text{Stokes I} \\
\frac{37}{10} \kappa^{-8/5}, &\text{Stokes Q} \\
\frac{13}{5} \kappa^{- (6/5)^{2}}, &\text{Stokes V}
\end{cases}
\end{equation}
are the best fit for $3 \leq \kappa \leq 7$.

We now turn to the absorptivity.  The low frequency fit is 
\begin{multline}
A_{S,lo} = X^{-2/3}_\kappa 3^{1/6} \frac{10}{41} \frac{(2 \pi)}{(w \kappa)^{10/3 - \kappa}} \frac{(\kappa-2)(\kappa-1)\kappa}{3\kappa-1} \Gamma(\frac{5}{3})  _{2}F_{1}(\kappa - \frac{1}{3}, \kappa + 1, \kappa + \frac{2}{3}, - \kappa w) \\
\, \times \, 
\begin{cases}
1 & \text{Stokes I} \\
-\frac{25}{48} & \text{Stokes Q} \\
0 &\text{Stokes U} \\
-\frac{77}{100 w}\left[(\sin{\theta})^{-114/50} -1\right]^{223/500} X_{\kappa}^{-7/20}\kappa^{-7/10} &\text{Stokes V}
\end{cases} \qquad
\end{multline}
where $_{2}F_1$ is a hypergeometric function.
For high frequency
\begin{multline}
A_{S,hi} = X^{-(1+\kappa)/2}_\kappa
\frac{\pi^{3/2}}{3} \frac{(\kappa-2)(\kappa-1)\kappa}{(w \kappa)^3}  (\frac{2 \Gamma(2+\kappa/2)}{2+\kappa}-1) \\
\, \times \, 
\begin{cases}
\left((\frac{3}{\kappa})^{19/4}+\frac{3}{5}\right) & \text{Stokes I} \\
-(21^{2} \kappa^{- (\frac{12}{5})^{2}} + \frac{11}{20}) & \text{Stokes Q} \\
0 &\text{Stokes U} \\
-\frac{143}{10}w^{-\frac{116}{125}}\left[(\sin{\theta})^{-\frac{41}{20}} -1\right]^{1/2}(13^{2}\kappa^{-8}+\frac{13}{2500}\kappa - \frac{1}{200}+\frac{47}{200\kappa})X_{\kappa}^{-1/2} &\text{Stokes V}
\end{cases} \qquad
\end{multline}
The final approximation is obtained by taking
\begin{equation}
A_S = \left(A_{S,lo}^{-x}+ A_{S,hi}^{-x}\right)^{-1/x},
\end{equation}
where
\begin{equation}
x = 
\begin{cases}
(- \frac{7}{4} + \frac{8}{5} \kappa)^{-43/50}, & \text{Stokes I} \\
\frac{7}{5} \kappa^{- 23/20}, & \text{Stokes Q} \\
\frac{61}{50}\kappa^{-142/125} + \frac{7}{1000}, & \text{Stokes V}.
\end{cases}
\end{equation}

\subsubsection{Fit errors}

\begin{table}[ht]
\centering
\begin{tabular}{ccccccccc}
\hline\hline
Stokes & emission & absorption & $\nu/\nu_{c}$ & $\theta$ & $\Theta_e$ & $p$ & $\kappa$ & $w$\\
Parameter &  &  &  & deg\\
\hline
\multicolumn{6}{c}{Thermal Distribution} \\
\hline
I & 35\% & 35\% & 10 & 15  & 3 & --- & --- & ---\\
Q & 5\% & 5\%   & 10 & 15  & 3 & --- & --- & ---\\
V & 50\% & 5\%  & 10 & 15  & 3 & --- & --- & ---\\
\hline
\multicolumn{6}{c}{Power-Law Distribution} \\
\hline
I & 35\% & 25\% & 10 & 15 & --- & 1.5 & --- & ---\\
Q & 20\% & 25\% & 10 & 15 & --- & 1.5 & --- & ---\\
V & 25\% & 30\% & 10 & 15 & --- & 1.5 & --- & ---\\
\hline
\multicolumn{6}{c}{Kappa Distribution} \\
\hline
I & 35\% & 40\% & 10 & 15 & --- & --- & 2.5 & 3\\
Q & 15\% & 35\% & 10 & 15 & --- & --- & 2.5 & 3\\
V & 25\% & 60\% & 10 & 15 & --- & --- & 2.5 & 3\\
\hline
\end{tabular}
\caption{Maximum Relative Errors in the fitting formulae and the parameters at which they occur.}
\label{tab:fiterr}
\end{table}

Table \ref{tab:fiterr} lists the maximum relative errors for each of the fitting functions along with the parameters where the maximum error occurs.
In all cases we have checked the fitting formulae over $10 < \nu/\nu_c < 3 \times 10^{10}$ and $15$ deg $< \theta < 85$ deg. The range of distribution specific parameters are $3 < \Theta_e < 40$ for the thermal distribution, $1.5 < p < 6.5$ for the power-law distribution, and $3 < w < 40$, $2.5 < \kappa < 7.5$ for the kappa distribution. The fits are accurate except for the lower and upper edges of the parameter regimes for $\nu/\nu_c$, $\theta$, as well as the distribution specific parameters, as can been seen in table \ref{tab:fiterr}.

\section{Conclusion}\label{sec:conclusion}

We have provided a code, {\tt symphony}, to calculate emissivities and absorptivities for arbitrary gyrotropic electron distribution functions.  The latest version of the code is available on github.\footnote{for the current version see {\tt http://github.com/afd-illinois/symphony}}  Along the way we have also provided fitting formulae for three distribution functions: thermal, power-law, and kappa.  

In {\tt symphony} the Stokes parameter is fixed as a parameter, as is the distribution function.  The emissivity is returned by a function {\tt j\_nu()}, which takes the electron number density, the magnetic field strength, frequency, and observer angle $\theta$ as arguments.  A similar function {\tt alpha\_nu()} returns the absorptivity.  Code to evaluate the fitting formulae is embedded in the full {\tt symphony} code and can be called using {\tt j\_nu\_fit()} and {\tt alpha\_nu\_fit()}.

Our investigation has uncovered a few interesting points.  First, the source function for a kappa distribution function is thermal ($\propto \nu^2$) at low frequency and nonthermal ($\propto \nu^{5/2}$) only at high frequency.  Second, all models produce nonnegligible circular polarization at $\nu/\nu_c \lesssim 10^3$ for our standard distribution function parameters. Third, our procedure permits a direct  evaluation of the accuracy of classic asymptotic formulae for the power-law distribution function; these formulae are good to better than a percent for $\nu/(\nu_c \sin\theta) \gtrsim 10^{2.5}$ for the $p = 3$ distribution considered here.  Fourth, the linear polarization fraction for a thermal distribution function approaches 1 for $\nu \gg \Theta_e^2 \nu_c$, and a similar result should obtain above any exponential cutoff in the distribution function.

Our codes may be useful to those seeking to understand the polarization properties of synchrotron emitting plasmas.  All results depend on observer angle $\theta$ (angle between photon and the magnetic field) but in some circumstances it may be useful to average over direction if the field is weak and tangled, or if the field has significant unresolved structure driven by Larmor-scale instabilities.  

\acknowledgements
This work was supported by NSF grant AST-1333612, a Lorella M. Jones undergraduate Summer Research Award to ZZ, a Simons Fellow in Theoretical Physics award and an All Souls College, Oxford Visiting Fellowship to CG, an Illinois Distinguished Fellowship and a University of California, Berkeley Visiting Scholar appointment to MC.

\end{document}